\documentclass[11pt,twoside]{article}
\usepackage{asp2010}
\usepackage{graphics}
\resetcounters

\begin{document}
\aspcpryear{2010}
\aspvoltitle{UP: Have Observations Revealed a Variable Upper End of
  the Initial Mass Function?}

\title{The galaxy-wide IMF - from star clusters to galaxies}
\author{Carsten~Weidner$^1$, Jan~Pflamm-Altenburg$^2$, and Pavel~Kroupa$^2$
\affil{$^1$SUPA, School of Physics \& Astronomy, University of
  St Andrews, North Haugh, St Andrews, Fife KY16 9SS, UK}
\affil{$^2$Argelander-Institut f\"ur Astronomie (Sternwarte), Auf dem
  H{\"u}gel 71, D-53121 Bonn, Germany}}

\begin{abstract}
Over the past years observations of young and populous star clusters
have shown that the stellar initial mass function (IMF) can be
conveniently described by a two-part power-law with an exponent
$\alpha_2$  = 2.3 for stars more massive than about 0.5 $M_\odot$ and
an exponent of $\alpha_1$ = 1.3 for less massive stars. A
consensus has also emerged that most, if not all, stars form in
stellar groups and 
star clusters, and that the mass function of these can
be described as a power-law (the embedded cluster mass function, ECMF)
with an exponent $\beta$ $\approx$ 2. These two results imply that the
integrated galactic IMF (IGIMF) for early-type stars cannot be a
Salpeter power-law, but that they must have a steeper exponent. An
application to star-burst galaxies shows that the IGIMF 
can become top-heavy. This has important consequences for the
distribution of stellar remnants and for the chemo-dynamical and
photometric evolution of galaxies. In this contribution the IGIMF
theory is described, and the accompanying contribution by
Pflamm-Altenburg, Weidner \& Kroupa (this volume) documents the
applications of the IGIMF theory to galactic astrophysics.
\end{abstract}

\section{Intro}
The stellar initial mass function (IMF) defines the ratio of
low-mass stars ($<$ 1 $M_\odot$), which do not contribute to the
chemical evolution over a Hubble time but lock-up baryonic matter, to
high-mass stars ($>$ few $M_\odot$), which power the interstellar
medium and enrich it 
with metals through AGB-winds and supernovae. It
further determines the mass-to-light ratios of stellar populations and
influences the dynamical evolution of star clusters and whole galaxies.

Over the last years it had become clear that star formation takes place
mostly in embedded clusterings \citep{LL03,AMG07}, each cluster or group
containing a dozen to many millions of stars \citep{Kr04b}. Within
these clusters stars appear to form following the canonical IMF,
$\xi(m) \propto m^{-\alpha}$, with a slope of 1.3 for stars with
$m~\le$ 0.5 $M_\odot$ and the Salpeter/Massey-slope of 2.35 for $m~>$
0.5 $M_\odot$ stars.

A result of clustered star-formation is that the composite stellar
population in a galaxy, which results from many star-forming events,
is the sum of the dissolving star clusters. Thus the integrated
galactic initial mass function (IGIMF) is the sum of all the IMFs of
all the star clusters \citep{KW03,WK05a}.

But (young, embedded) star clusters also follow a mass
function. The embedded cluster mass function (ECMF) has been found to 
be a power-law, $\xi_\mathrm{ecl} \propto M_\mathrm{ecl}^{-\beta}$, with
a rather constant slope of $\approx 2$ for largely different
environments from the quiescent solar neighbourhood to the vigorously
star-forming Antennae galaxies \citep{LL03,HEDM03,ZF99}.

Additionally, it appears that star clusters limit the mass of the
most-massive star that can form within them. Low-mass clusters are
unable to form very massive stars. This follows not only from
observations but is a necessary logical statement given that a finite
mass reservoir is distributed over the stellar population with an
invariant IMF shape. This results in a relation between cluster mass
and the most-massive star \citep{WK04,WK05b,WKB09}.

A direct consequence is that the IGIMF is steeper than the individual
canonical IMFs in the actual clusters, hereby immediately explaining
why $\alpha_{3, \mathrm{field}} = 2.7 > \alpha_3 = 2.35$, where 
$\alpha_{3,\mathrm{field}}$ is the slope of the IMF derived by
\citet{Sc86} and \citet{RGH02} from OB star counts in the Milky Way
field, and $\alpha_3$ = 2.35 is the \citet{Sal55} index. This is due
to the fact that low-mass star clusters are numerous but can not have
massive stars.

A mathematical description of how the IGIMF\footnote{A tool
to calculate an IGIMF and to fit it with a multi-part power law as
required as the input IMF for P{\sc egase} has been presented in
\citet{PWK09} and can be downloaded from
\tt{www.astro.uni-bonn.de}} can be calculated is given 
in \S~\ref{sec:math}, while \S~\ref{sec:cons} will discuss
observational consequences of the IGIMF for the stellar populations of
galaxies.

\section{Mathematical formulation of the IGIMF}
\label{sec:math}
In order to describe the IGIMF mathematically, three ingredients are
necessary. The IMF within star clusters (described in
\S~\ref{sub:IMF}), how the cluster mass limits its most-massive star
(\S~\ref{sub:mmax}) and how the cluster mass function (ECMF) is
defined by the star-formation rate (SFR) of a galaxy
(\S~\ref{sub:SFR}). Table \ref{tab:ref} gives an overview of the
step-by-step development of the IGIMF theory.

\subsection{The stellar IMF in star clusters}
\label{sub:IMF}
An arbitrary multi-power law distribution function is
{\small
\begin{equation}
\xi(m) = k \left\{\begin{array}{ll}
k^{'}\left(\frac{m}{m_{\rm H}} \right)^{-\alpha_{0}}&\hspace{-0.25cm},m_{\rm
  low} \le m < m_{\rm H},\\
\left(\frac{m}{m_{\rm H}} \right)^{-\alpha_{1}}&\hspace{-0.25cm},m_{\rm
  H} \le m < m_{0},\\
\left(\frac{m_{0}}{m_{\rm H}} \right)^{-\alpha_{1}}
  \left(\frac{m}{m_{0}} \right)^{-\alpha_{2}}&\hspace{-0.25cm},m_{0}
  \le m < m_{1},\\ 
\left(\frac{m_{0}}{m_{\rm H}} \right)^{-\alpha_{1}}
    \left(\frac{m_{1}}{m_{0}} \right)^{-\alpha_{2}}
    \left(\frac{m}{m_{1}} \right)^{-\alpha_{3}}&\hspace{-0.25cm},m_{1}
    \le m < m_{\rm max}.\\ 
\end{array} \right. 
\label{eq:4pow}
\end{equation}
\noindent The IMF in star clusters has been found to be conveniently
described by the following set of indices.
\begin{equation}
          \begin{array}{l@{\quad\quad,\quad}l}
\alpha_0 = +0.30&0.01 \le m/{M}_\odot < 0.08,\\
\alpha_1 = +1.30&0.08 \le m/{M}_\odot < 0.50,\\
\alpha_2 = +2.30&0.50 \le m/{M}_\odot  \le m_\mathrm{max}(M_\mathrm{ecl}),\\
          \end{array}
\label{eq:imf}
\end{equation}}
\noindent where $dN = \xi(m)\,dm$ is the number of stars in the mass
interval $m$ to $m + dm$ and $m_\mathrm{max}$ is the mass of the
most-massive star in a cluster which is regulated by the cluster
mass, $M_\mathrm{ecl}$. The exponents $\alpha_{\rm i}$ represent the
standard or canonical IMF \citep{Kr01,Kr02}, and $k^{'}$ signifies the
discontinuity near the brown dwarf / stellar mass limit
\citep{TK07}. Note that $\alpha_3~\approx~\alpha_2$ so that the
canonical IMF is excellently described by a very convenient two-part
power law distribution function.

\subsection{The most-massive star in a star cluster}
\label{sub:mmax}

\begin{figure}
\plottwo{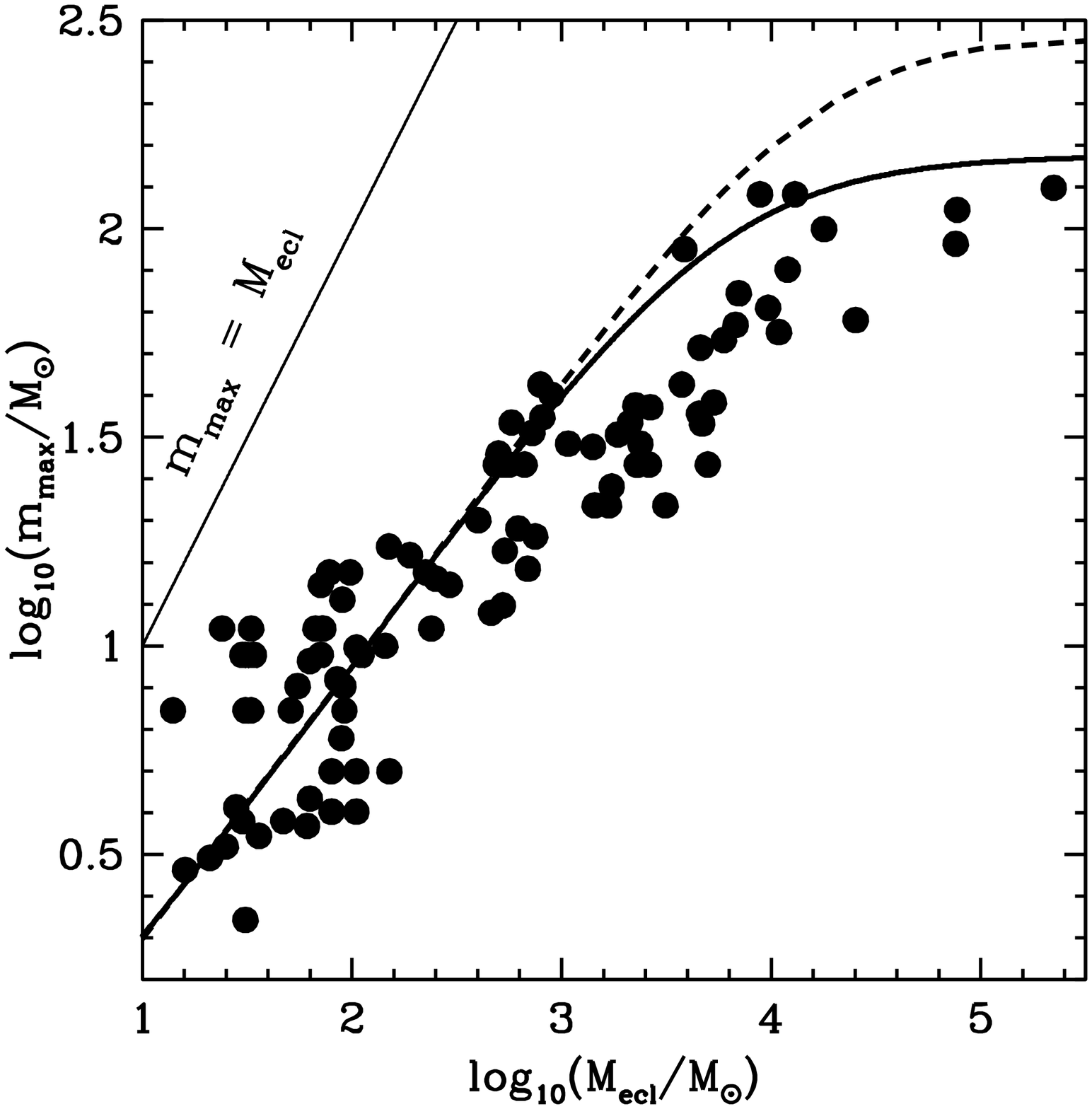}{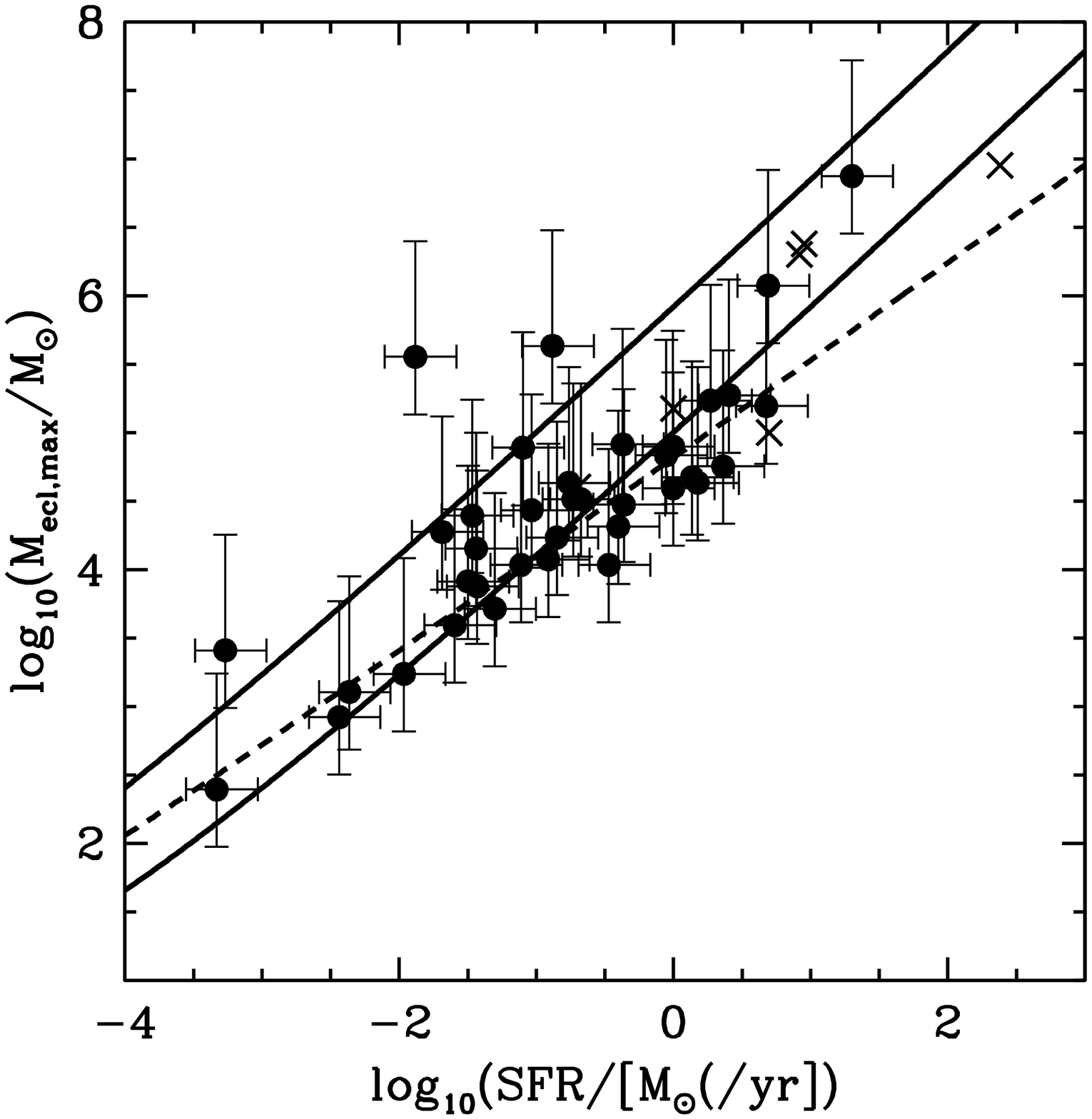}
\vspace*{-1cm}
\caption{Left: The most-massive star, $m_{\rm max}$, vs the embedded cluster
  mass, $M_{\rm ecl}$. The dots are observations from \citet{WKB09}
  and the solid line is a numerical solution to eqs.~\ref{eq:Mecl} and
  \ref{eq:normlim} assuming $m_\mathrm{max*}$ = 150 $M_\odot$, while
  for the dashed line $m_\mathrm{max*}$ is 300 $M_\odot$. Right:
  Maximum cluster mass versus global star-formation rate (SFR), 
both in logarithmic units. Filled dots are observations compiled by
\citet{WKL04} with error estimates and the crosses are additional
recent data points. The top solid line is the result of
eqs.~\ref{mtotb} and \ref{mtotc} with $\beta$ = 2.0 and $\delta t$ =
10 Myr and the bottom solid line of $\beta$ = 2.0 and $\delta t$ =
1 Myr, while the dashed line assumes $\beta$ = 2.4 and $\delta t$ =
10 Myr.
\label{fig:mmax}
}
\end{figure}

In order to analytically determine the mass of the most-massive star
in a cluster, the fact that there must be one most-massive star is
used,
\begin{equation}
\label{eq:normlim}
1 = \int_{m_{\rm max}}^{m_{\rm max *}}\xi(m)~dm,
\end{equation}
where $m_\mathrm{max *}$ is the fundamental upper mass limit for stars
which was found to be $m_\mathrm{max *}$ $\approx$ 150 $M_\odot$
\citep{WK04,Fi05,Ko06}, though new indications exists that it may be as
high as 300 $M_\odot$ \citep{CSH10}. 

After inserting eq.~\ref{eq:4pow} into eq.~\ref{eq:normlim} an equation
with two unknowns, $k$ and $m_\mathrm{max}$, emerges. Therefore, an
additional constraint is necessary in order to solve these
equations. Such a constraint is given by the mass of the cluster,
$M_\mathrm{ecl}$.

\begin{equation}
\label{eq:Mecl}
M_{\rm ecl} = \int_{m_{\rm low}}^{m_{\rm max}}m \cdot \xi(m)~dm.
\end{equation}

\noindent The low mass end of the IMF, $m_\mathrm{low}$, is set to 0.01 
$M_\odot$. Eqs.~\ref{eq:normlim} and \ref{eq:Mecl} can not be solved
explicitly but the existence of a unique solution can be proven \citep{PAK06}.

The left panel of Fig.~\ref{fig:mmax} shows the numerical solution of
the system of equations as a solid (with $m_\mathrm{max *}$ = 150
$M_\odot$) and a dashed line ($m_\mathrm{max *}$ = 300 $M_\odot$)
together with observations of most-massive stars in star clusters
\citep{WK04,WK05b,WKB09}. The observations in the left panel of
Fig.~\ref{fig:mmax} are also an indication that stars are not randomly
sampled from the IMF in a star cluster \citep{WK05b,HA10}. Therefore, 
physical processes might link the formation of the most-massive star
to the potential of the proto-cluster cloud core. 

\subsection{The relation between the SFR and the most-massive star
  cluster}
\label{sub:SFR}

In order to calculate the IGIMF for a galaxy it is needed to know how
a star cluster population is build-up in a galaxy. 

The aim now is to estimate the star-formation rate (SFR) required to
build a complete young star-cluster population in one star-formation
epoch such that it is populated fully with masses ranging up to
$M_{\rm ecl,max}$. Observational surveys suggest the embedded-cluster mass
function (ECMF) is a power-law,
\begin{equation} 
\label{eq:cmf}
\xi_{\rm ecl}(M_{\rm ecl}) = k_{\rm ecl} \cdot
  \left(\frac{M_{\rm ecl}}{M_{\rm ecl,max}}\right)^{-\beta},
\end{equation}
with $\beta$ $\approx$ 2 \citep{ElEf97,Kr02,KB02,LL03}. For the total
mass of a population of young stellar clusters formed in a time-span
$\delta t$,
\begin{equation} 
\label{Mint}
M_{\rm tot} = \int_{M_{\rm ecl,min}}^{M_{\rm ecl,max}}M_{\rm ecl}
  \cdot \xi_{\rm ecl}(M_{\rm ecl})~dM_{\rm ecl},
\end{equation}
where $M_{\rm ecl,max}$ is the mass of the heaviest cluster in the
population.  The normalisation constant $k_{\rm ecl}$ is determined by
stating that $M_{\rm ecl,max}$ is the single most massive cluster,
\begin{equation} 
\label{Nint}
1 = \int_{M_{\rm ecl,max}}^{\infty} \xi_{\rm ecl}(M_{\rm ecl})~dM_{\rm ecl}.
\end{equation}

\noindent With an ECMF power-law index of $\beta = 2$ we get from
eq.~\ref{Nint}, 
\begin{equation} \label{norm} 
k_{\rm ecl} = \frac{1}{M_{\rm ecl,max}}.
\end{equation}
Inserting this into eq.~\ref{Mint} (again with $\beta$ = 2),
\begin{equation} \label{mtot} M_{\rm tot} = M_{\rm ecl,max} \cdot
  (\ln{M_{\rm ecl,max}}-\ln{M_{\rm ecl,min}}).
\end{equation}
$M_{\rm ecl,min}$ is the minimal cluster mass which we take to be
$5\,M_{\odot}$ (a small Taurus-Auriga like sub-group). For arbitrary
$\beta \ne 2$ eqs.~\ref{norm} and \ref{mtot} change to
\begin{equation} \label{normb} k_{\rm ecl} = \frac{\beta-1}{M_{\rm
      ecl,max}}
\end{equation}
and
\begin{equation} \label{mtotb} M_{\rm tot} = (\beta-1) \cdot M_{\rm
    ecl,max}^{\beta-1} \cdot \left(\frac{M_{\rm ecl,max}^{2-\beta}-M_{\rm
    ecl,min}^{2-\beta }}{2-\beta}\right).
\end{equation}

Given a SFR, a fully-populated ECMF with total mass $M_{\rm tot}$ is
constructed in the time $\delta t$,
\begin{equation} \label{mtotc} M_{\rm tot} = SFR \cdot \delta t.
\end{equation}

Thus, dividing $M_{\rm tot}$ by a formation time,
$\delta t$, of 10 Myr, and varying $M_{\rm ecl,max}$ between 10$^1$
and 10$^8$ $M_\odot$, results in a theoretical $M_{\rm ecl,max}(SFR)$
relation which is shown as solid lines ($\beta$ = 2 and $\delta t$ =
10 and 1 Myr) and a dashed line ($\beta$ = 2.4 and $\delta t$ = 10
Myr) in the right panel of Fig.~\ref{fig:mmax} \citep{WKL04}. Note
that significantly other values of $\delta t$ lead to a wrong slope
and a wrong normalisation of the $M_\mathrm{ecl,max}$ = fn(SFR)
relation if $\beta$ = 2.4. For example, for $\beta$ = 2, a $\delta t$
of 1 Myr is necessary for the relation to reproduce the data.

\subsection{The IGIMF}
\label{sub:IGIMF}
\begin{figure}
\centering
\includegraphics[scale=.80]{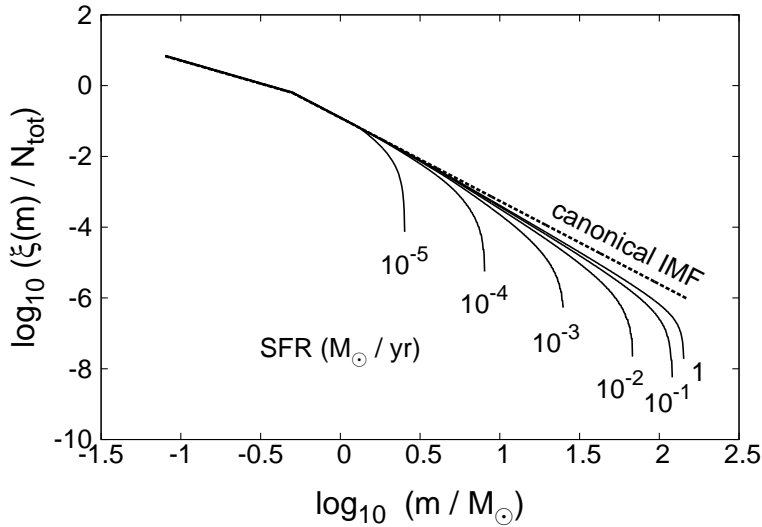}
\caption{The IGIMF for different SFRs in $M_\odot$/yr. Each IGIMF is normalised
such that $\int \xi_\mathrm{IGIMF}(m)\;dm=1$.
\label{fig:IGIMF}
}
\end{figure}

With the IMF and the two so far derived relations, the
$M_\mathrm{ecl}$-$m_\mathrm{max}$-, and the SFR-$M_\mathrm{ecl,
max}$-relation, it is now possible to formulate and calculate the
IGIMF as follows,

\begin{equation} 
\label{eq:igimf}
\xi_{\rm IGIMF}(m,t) = \int_{M_{\rm ecl,min}}^{M_{\rm ecl,max}(SFR(t))}
\xi(m\le m_{\rm max}(M_\mathrm{ecl}))~\xi_{\rm ecl}(M_{\rm ecl})~dM_{\rm ecl}. 
\end{equation}
Thus $\xi(m\le m_{\rm max})~\xi_{\rm ecl}(M_{\rm ecl})~dM_{\rm ecl}$ 
is the stellar IMF, with $m_\mathrm{max}$ limited by $M_\mathrm{ecl}$, 
contributed by $\xi_{\rm ecl}~dM_{\rm ecl}$ clusters with mass near
$M_{\rm ecl}$. While $M_{\rm ecl,max}$ follows from the
SFR-$M_\mathrm{ecl,max}$-relation, $M_{\rm ecl,min}\,=\,5\,M_{\odot}$ 
is adopted. The resulting IGIMF for a number of SFRs is shown in
Fig.~\ref{fig:IGIMF} \citep{KW03,WK05a,WK05b,PWK07}. Recently, the
IGIMF has also been applied to starbursts \citep{WK10}, and is briefly
touched upon below in Fig.~\ref{fig:top}. 

\begin{table}
\caption{\label{tab:ref} IGIMF-ingredients}
\begin{tabular}{c|c}
Ingredient&Reference\\
\hline
basic formulation&\citet{KW03}\\
$m_\mathrm{max}$-$M_\mathrm{ecl}$-relation&\citet{WK04,WK05b,WKB09}\\
SFR-$M_\mathrm{ecl,max}$-relation&\citet{WKL04}\\
variation with SFR&\citet{WK05a,PWK07}\\
application to starbursts&\citet{WK10}\\
\end{tabular}
\end{table}

\section{Consequences of the IGIMF}
\label{sec:cons}
Physically, the IGIMF follows from a few natural assumptions: stars do
not form in isolation but in groups and clusters and star-formation is
not purely stochastic but processes like stellar feedback regulate
star-formation and therefore link their environment (cluster mass)
with the final product (most-massive star). 

Mathematically, the set of equations developed above from first
principles lead to a remarkably successful description of star formation
in clusters and galaxies.

It can be seen in Fig.~\ref{fig:IGIMF} that the IGIMF is generally
steeper than the IMF seen in individual star clusters. This
has profound consequences for galaxies. \citet{KW03} showed that the
supernova type II rate can be influenced strongly by the IGIMF. This
has a significant impact on the chemical evolution of galaxies and has
been studied by \citet{KWK05} who show that the mass-metallicity
relation resulting from the IGIMF agrees well with
observational results. Other studies arrive at similar results
\citep{RCK09}. With the IGIMF the mass-metallicity properties of
galaxies thus emerge very naturally.

\begin{figure}[t]
\centering
\includegraphics[scale=.4]{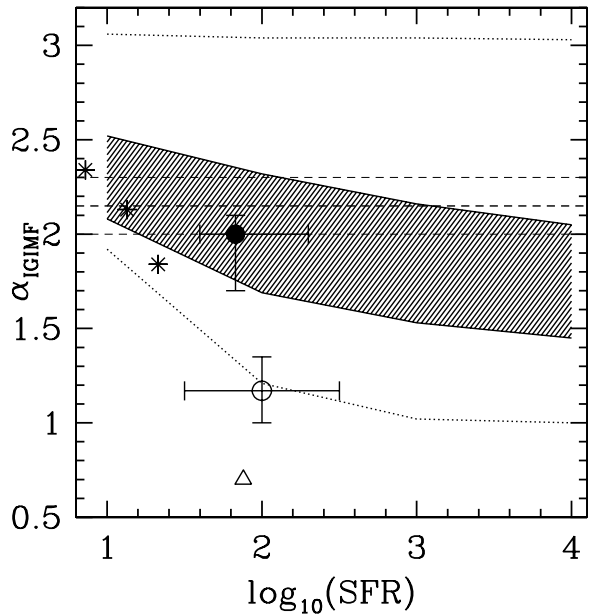}
\vspace*{-1.5cm}
\caption{IGIMF slopes above 1 $M_\odot$ ($\alpha_\mathrm{IGIMF}$) when
assuming top-heavy starburst clusters and some observational
constrains in dependence of the total SFR. The shaded region
  between the solid lines marks the range of model results for
  an ECMF slope $\beta$ = 2.0 while the dotted lines show the
  full envelope constrained by all models. The three asterisks
  are from \citet{Da07}, the one triangle around $\alpha_3
  \approx 0.7$ is from \citet{VD07}, the filled circle with
  error bars is from \citet{BKM07} while the open circle with
  error bars corresponds to the \citet{BLF05} result. The dashed
  horizontal line marks the \citet{WHT08} constrain with the light
  dashed lines 0.15 dex above and below being their uncertainty
  range.
\label{fig:top}
}
\end{figure}

As is shown in another contribution in these proceedings
(Pflamm-Altenburg et al.), the IGIMF predicts a discrepancy
between SFRs measured by UV and by H$\alpha$ \citep{PWK07,PWK09} again
naturally explaining recent observational claims of such a difference 
\citep{MWK09,LGT09}. In addition to these results it is possible to
expand the IGIMF theory towards local rather than global
properties. This local IGIMF (LIGIMF) explains well the observed
radial H$\alpha$ cut-off in disk galaxies \citep{PAK08}. 

Furthermore, it can be shown that the IGIMF theory does not
necessarily only produce steep (bottom heavy) galaxy-wide IMFs but
also top-heavy IGIMFs are possible: When assuming that at very high
SFRs extremely massive star clusters are formed it has been shown that
these will most likely produce top-heavy IMFs. As star clusters are
generally of similar physical size regardless of their mass
\citep[$r_{\rm ecl}\,\lesssim\,1$~pc,][]{TPN98,Kr04b,GMP05,RJS06,SHG07}, 
and as proto-stars are much larger than main-sequence stars
\citep[with radii between 5000 and 20000 AU,][]{BAW01,FKS06,vdT00},
crowding of the proto-stars can happen and might change the slope of
the high-mass IMF in such clusters \citep{BBZ98,ES03,Sh04,MKD10}. If
sufficient quantities of such massive star clusters are formed,
e.g.~in a starburst, the whole IGIMF can become top-heavy
\citep{WK10}. The resulting IGIMFs slopes above 1 $M_\odot$ for
top-heavy IMFs in starbursts are shown in Fig.~\ref{fig:top} together
with various observational estimates. Indeed, in order to describe the
cosmological evolution of stellar mass in galaxies a top-heavy IGIMF
seems to be needed \citep{WHT08,WTH08}.

\bibliography{WPK_UP2010}

\end{document}